\documentclass[aps, prd, amsmath, floats, floatfix, twocolumn, preprintnumbers, superscriptaddress, nofootinbib, showpacs, eqsecnum]{revtex4-1}
\usepackage{graphicx}
\usepackage{amssymb}
\usepackage{color}

\newcommand{\be}{\begin{eqnarray}}
\newcommand{\ee}{\end{eqnarray}}

\begin{document}

\title{Direct imaging rapidly-rotating non-Kerr black holes}

\author{Cosimo Bambi}
\email{Cosimo.Bambi@physik.uni-muenchen.de}
\affiliation{Arnold Sommerfeld Center for Theoretical Physics,
Ludwig-Maximilians-Universit\"at M\"unchen, 80333 Munich, Germany}

\author{Francesco Caravelli}
\email{fcaravelli@perimeterinstitute.ca}
\affiliation{Max Planck Institute for Gravitational Physics, 
Albert Einstein Institute, 14476 Golm, Germany}
\affiliation{Department of Physics, University of Waterloo, 
Waterloo, Ontario N2L 3G1, Canada}
\affiliation{Perimeter Institute for Theoretical Physics, 
Waterloo, Ontario N2L 2Y5, Canada}

\author{Leonardo Modesto}
\email{lmodesto@perimeterinstitute.ca}
\affiliation{Perimeter Institute for Theoretical Physics, 
Waterloo, Ontario N2L 2Y5, Canada}

\date{\today}

\begin{abstract}
Recently, two of us have argued that non-Kerr black holes in 
gravity theories different from General Relativity may have a 
topologically non-trivial event horizon. More precisely, the 
spatial topology of the horizon of non-rotating and slow-rotating 
objects would be a 2-sphere, like in Kerr space-time, while 
it would change above a critical value of the spin parameter. 
When the topology of the horizon changes, the black hole central
singularity shows up. The accretion process from a thin disk 
can potentially overspin these black holes and induce the topology
transition, violating the Weak Cosmic Censorship Conjecture. If 
the astrophysical black hole candidates are not the black holes
predicted by General Relativity, we might have the quite unique 
opportunity to see their central region, where classical physics 
breaks down and quantum gravity effects should appear. Even if the 
quantum gravity region turned out to be extremely small, at the 
level of the Planck scale, the size of its apparent image would be 
finite and potentially observable with future facilities.
\end{abstract}

\maketitle


\section{Introduction}

General Relativity (GR) is our current theory of gravity and up to now 
there is no clear observational evidence in disagreement with
its predictions. Nevertheless, the theory has been tested only
for weak gravitational fields, where $g_{tt} \approx - (1 + \phi)$ and 
$|\phi| \ll 1$~\cite{will}. There are instead physically interesting 
situations in which GR breaks down and predicts 
space-time singularities and regions with closed time-like curves. 
Here, (unknown) quantum gravity effects are thought to become 
important. According to the Weak Cosmic Censorship Conjecture 
(WCCC)~\cite{ccc}, these quantum gravity regions are always 
hidden behind an event horizon and we would have no chances 
to observe them. However, the assumption of the WCCC is 
essentially motivated to assure the validity of classical gravity 
in any astrophysical situation, with the guess that new physics 
can never be seen by a distant observer. For the time being, 
there is no fundamental principle requiring that~\cite{accc}.

In 4-dimensional GR, uncharged rotating black holes (BHs) are described 
by the Kerr solution, which is completely specified by two parameters, 
the mass $M$ and the spin parameter $a$. The condition for the 
existence of the event horizon is $a \le M$, while for $a > M$ there 
is no BH but a naked singularity, which is forbidden by the 
WCCC\footnote{Throughout this paper, we use the convention 
$a = |a|$ and units in which $G_N = c = 1$.}. 
While it is not yet clear if naked singularities can be created in 
Nature~\cite{pankaj}, any attempt to make a star collapse with 
$a > M$~\cite{rezz}, as well as to overspin an already existing 
Kerr BH and get a naked singularity~\cite{enrico}, seems to be 
doomed to fail. In Ref.~\cite{pap1}, two of us considered the 
loop-inspired BHs proposed in~\cite{loopr} and the non-GR BHs 
introduced in~\cite{dim}. In addition to $M$ and $a$, these 
space-times have at least one more parameter: it can be seen as 
a ``deformation parameter'' and it produces deviations from the 
Kerr geometry. It was shown that the spatial topology of the horizon 
of these objects is a 2-sphere (like for a Kerr BH) in the non-rotating 
and slow-rotating case, while it changes above a critical value of 
$a/M$. Our conjecture is that such a phenomenon may be common for
non-Kerr BHs. The basic mechanism is the following. A Kerr BH with 
$a/M < 1$ has an outer horizon with radius $r_+$ and an inner 
horizon with radius $r_-$. As $a/M$ increases, $r_+$ decreases 
and $r_-$ increases. When $a/M = 1$, there is only one horizon 
(extremal black hole with $r_+ = r_-$) and, for $a/M > 1$, there
is no horizon. For the BHs studied in~\cite{pap1}, the outer and the
inner horizons have not the same shape. So, when $a/M$ increases, 
the two horizons still approach each other, but eventually merge 
forming a single horizon with non-trivial topology.
After the topology transition, the BH central singularity (or the high 
curvature region if the singularity is solved, as it occurs at least for
some loop-inspired BHs) is not hidden behind the horizon any more. 
Interestingly, such a rapidly rotating BHs can be easily
created: it is just necessary a thin accretion disk~\cite{pap1}.

\section{Black holes in alternative theories of gravity}

In this paper, we consider the BH metric proposed in
Ref.~\cite{dim}. However, as emphasized in~\cite{pap1},
the phenomenon we are going to study should depend only
marginally on this choice and be common to rapidly-rotating
non-Kerr BHs. Together with the loop-inspired BH
solution in~\cite{loopr}, they are the only known 4D 
electrically-neutral non-Kerr
metrics in analytical form valid beyond the non-rotating and 
slow-rotating approximation. The non-zero metric
coefficients, in Boyer-Lindquist coordinates, 
are~\cite{dim}\footnote{As suggested in Ref.~\cite{sup-ren},
a metric of this kind can be solution of a particular non-local 
generalization of the Einstein's equations:
$$  R_{\mu \nu} -\frac{1}{2} g_{\mu \nu} R = 
8 \pi G_N \mathcal{O}( \Box/\Lambda^2) T_{\mu \nu} \, ,$$ 
where $\mathcal{O}(\Box/\Lambda^2)$ is a generic non-local function 
of the covariant D'Alembertian operator and $\Lambda$ is the 
energy scale of the modified gravity.}
\be
g_{tt} &=& - \left(1 - \frac{2 M r}{\rho^2}\right) (1 + h)
 \, , \nonumber\\
g_{t\phi} &=& - \frac{2 a M r \sin^2\theta}{\rho^2} 
(1 + h) \, , \nonumber\\
g_{\phi\phi} &=& \sin^2\theta \left[r^2 + a^2
+ \frac{2 a^2 M r \sin^2\theta}{\rho^2} \right] + \nonumber\\
&& + \frac{a^2 (\rho^2 + 2 M r) \sin^4\theta}{\rho^2} 
h \, , \nonumber\\
g_{rr} &=& \frac{\rho^2 (1 + h)}{\Delta + 
a^2 h \sin^2\theta } \, \,\,,\, \,\,\,
g_{\theta\theta} = \rho^2 \, ,
\label{eqm}
\ee
where
\be
\rho^2 &=& r^2 + a^2 \cos^2\theta \, , \nonumber\\
\Delta &=& r^2 - 2 M r + a^2 \, , \nonumber\\
h &=& \sum_{k = 0}^{\infty} \left(\epsilon_{2k} 
+ \frac{M r}{\rho^2} \epsilon_{2k+1} \right)
\left(\frac{M^2}{\rho^2}\right)^k \, .
\ee

Assuming that the function $h$ is infinitely time differentiable, 
the metric has an infinite number of free parameters $\epsilon_i$ 
and the Kerr solution is recovered when all these parameters 
are set to zero. As shown in~\cite{dim}, $\epsilon_0 = \epsilon_1 = 0$,
in order to recover the correct Newtonian limit, while Solar 
System experiments constrain $\epsilon_2$ at the level of 
$10^{-4}$. For the sake of simplicity, in the rest of the paper we 
restrict our attention to the case in which $\epsilon_3$ is the only 
deformation parameter and $\epsilon_i = 0$ for $i \neq 3$.
In this case, as discussed in~\cite{pap1}, current estimates of
the mean radiative efficiency of AGN seem to require 
$-1.1 < \epsilon_3 < 25$, but such a bound has to be 
taken with caution.

\begin{table}[b]
\begin{center}
\begin{tabular}{c c c c c}
\hline \\
\hspace{.5cm} & $R_{QG}/M$ & \hspace{.5cm} & $\Delta y/M$ &  \hspace{.5cm} \\ \\
\hline 
& $0.1$ & & 0.444 & \\
& $0.01$ & & 0.412 & \\ 
& $0.001$ & & 0.410 & \\
\hline
\end{tabular}
\end{center}
\caption{Vertical size of the primary image at $x = 0$ of the quantum 
gravity region of a black hole with $a/M = 1.18$ and $\epsilon_3 = -1.0$. 
The viewing angle of the distant observer is $i = 5^\circ$. $R_{QG}$ 
is the radius in Boyer-Lindquist coordinates of the quantum gravity 
region. See text for details.}
\label{tab}
\end{table}

\section{Direct imaging black holes with non-trivial topology \label{s-di}}

The geometry around a BH can be probed by observing the
direct image of its accretion flow, as already explored in~\cite{shadow}.
For a review on current and near future capabilities of Very Long
Baseline Interferometry (VLBI) experiments, see e.g.~\cite{vlbi} 
and references therein. If the gas of accretion is optically thin 
(which is always possible at sufficiently high frequencies) and 
geometrically thick, one sees the ``shadow''; that is, a dark area 
over a brighter background~\cite{rohta}. While the intensity map 
of the image strongly depends on the specific accretion model, 
i.e. on the configuration of the system and on complicated 
astrophysical processes, the contour of the shadow is determined 
only by the geometry of the space-time: basically, it is the photon 
capture surface as seen by a distant observer. For rotating BHs, 
the shape of the shadow is not symmetric with respect to the 
rotation axis, because the capture radius for corotating photons 
is smaller than the one for counterrotating photons. The effect is 
more evident for fast-rotating BHs and observers with a viewing angle 
$i = 90^\circ$, while it disappears for non-rotating BHs or observers 
along the rotation axis.

\begin{figure*}
\begin{center}  
\includegraphics[type=pdf,ext=.pdf,read=.pdf,width=5.2cm]{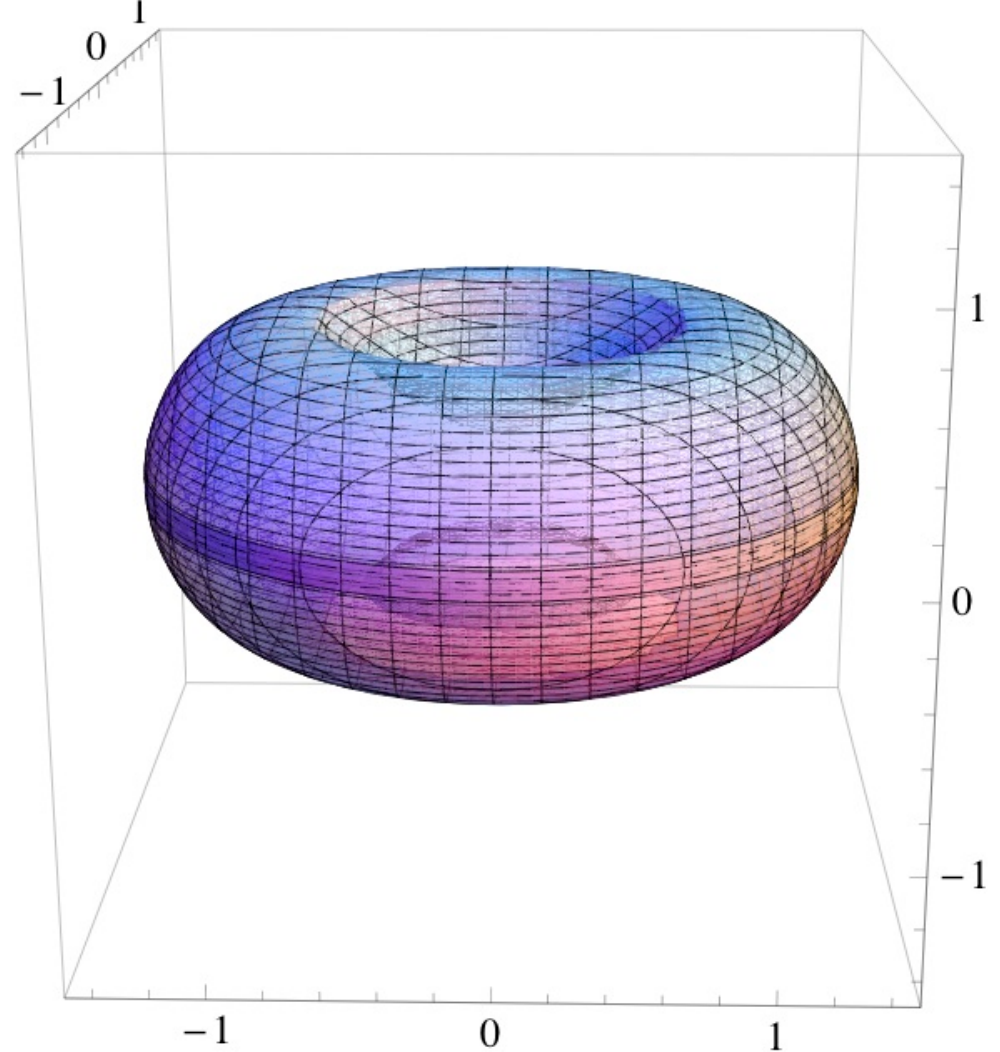}
\hspace{2.2cm}
\includegraphics[type=pdf,ext=.pdf,read=.pdf,width=5.5cm]{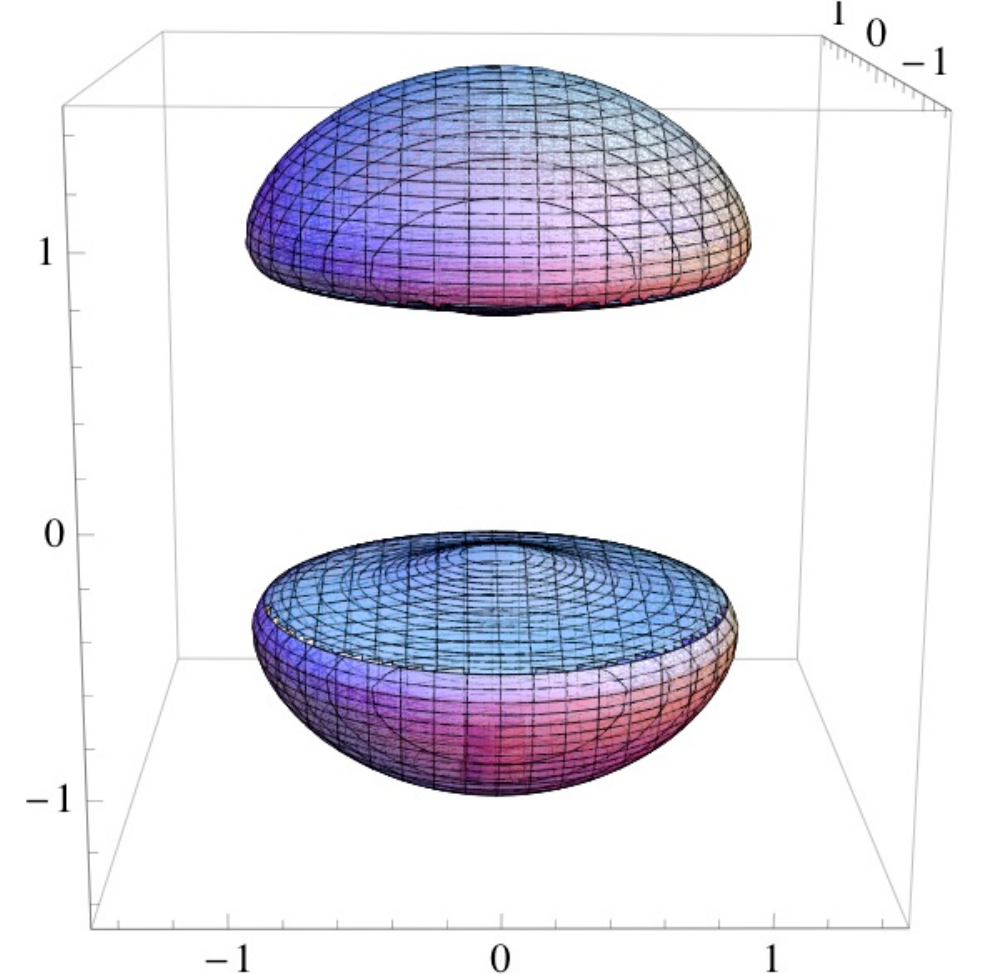}
 \end{center}
\caption{Event horizon of the two black holes studied in Sec.~\ref{s-di}.
Left panel: black hole with $a/M = 1.18$ and $\epsilon_3 = - 1.0$. 
Right panel: black hole with $a/M = 0.87$ and $\epsilon_3 = 1.0$.
Axes in units $M = 1$.}
\label{f4}
\end{figure*}

In this section, we consider two specific cases; this is enough to
illustrate the qualitative features of the BHs described by the 
metric~(\ref{eqm}). The first BH has $a/M = 1.18$ and $\epsilon_3 
= -1.0$. As $\epsilon_3 < 0$, the object is more oblate
than a Kerr BH. $a/M = 1.18$ corresponds to the equilibrium value
in the case the BH is accreting from a thin disk on the equatorial
plane, see Table~II in~\cite{pap1}. The equilibrium spin parameter
is indeed always larger than $M$ when a compact object is more
oblate than a Kerr BH~\cite{evol}. The horizon of this BH is defined
by $g^{rr} = 0$; that is:
\be\label{eqh}
\Delta + a^2 h \sin^2 \theta = 0
\ee
and it is shown in the left panel of Fig.~\ref{f4}. As already pointed out 
in Ref.~\cite{pap1}, rapidly-rotating oblate BHs have a shape similar 
to a donut. However, in the specific case of the metric~(\ref{eqm}),
there is no central hole: the event horizon extends up to $r = 0$,
where there is a naked singularity. As second example, we consider 
a BH with $a/M = 0.87$ and $\epsilon_3 = 1.0$. As in the previous case, 
this value of $a/M$ corresponds to the one of equilibrium when 
the BH is accreting from a thin disk on the equatorial plane. When
$\epsilon_3 > 0$, the object is more prolate than a Kerr BH and
its horizon, still given by Eq.~(\ref{eqh}), is formed by two disconnected
horizons, each of which with spatial topology of a 2-sphere, as
shown in the right panel of Fig.~\ref{f4}.

\begin{figure*}
\begin{center}  
\includegraphics[type=pdf,ext=.pdf,read=.pdf,width=5.5cm]{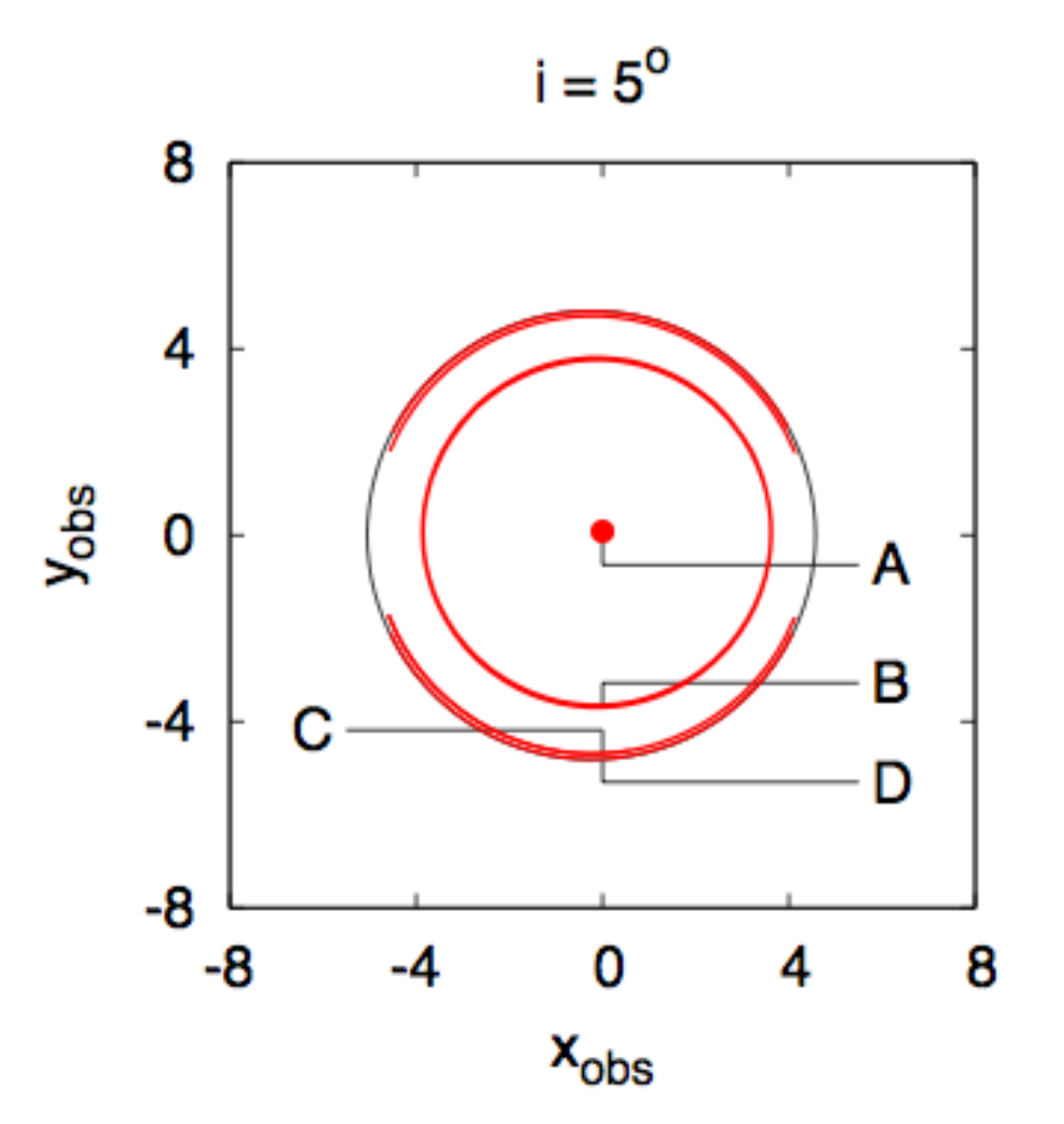}
\includegraphics[type=pdf,ext=.pdf,read=.pdf,width=5.5cm]{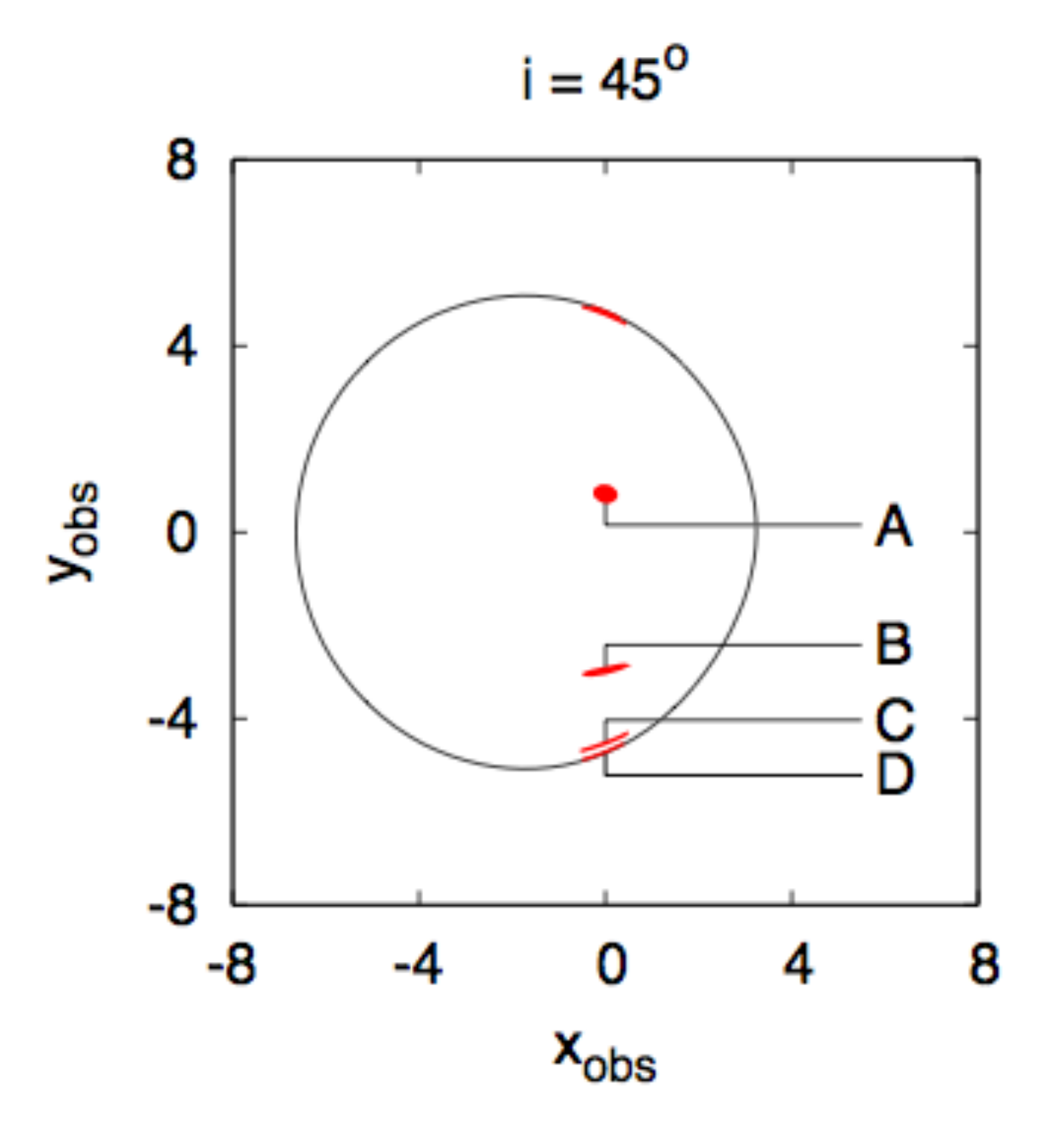}
\includegraphics[type=pdf,ext=.pdf,read=.pdf,width=5.5cm]{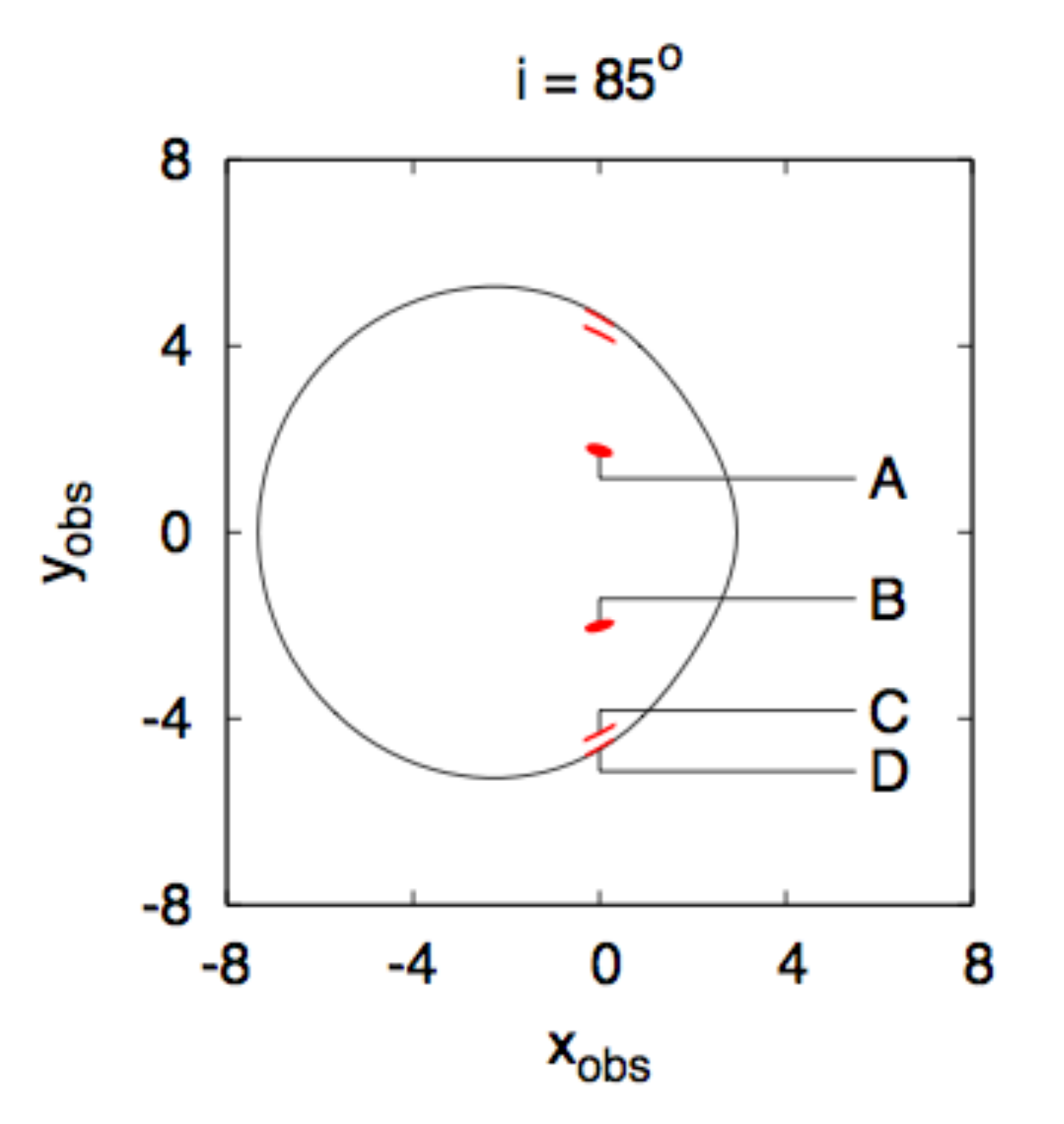}
 \end{center}
\vspace{-0.3cm}
\caption{Apparent image of the black hole with $a/M = 1.18$ and 
$\epsilon_3 = -1.0$ for a distant observer with viewing angle 
$i = 5^\circ$ (left panel), $45^\circ$ (central panel), and $85^\circ$ 
(right panel). The black curve is the contour of the black hole 
shadow. The red area is the image of the quantum gravity region.
The trajectories of the light rays A, B, C, and D around the black 
hole are shown in Fig.~\ref{f2}. Axes in units $M = 1$.}
\label{f1}
\vspace{0.4cm}
\begin{center}  
\includegraphics[type=pdf,ext=.pdf,read=.pdf,width=5.5cm]{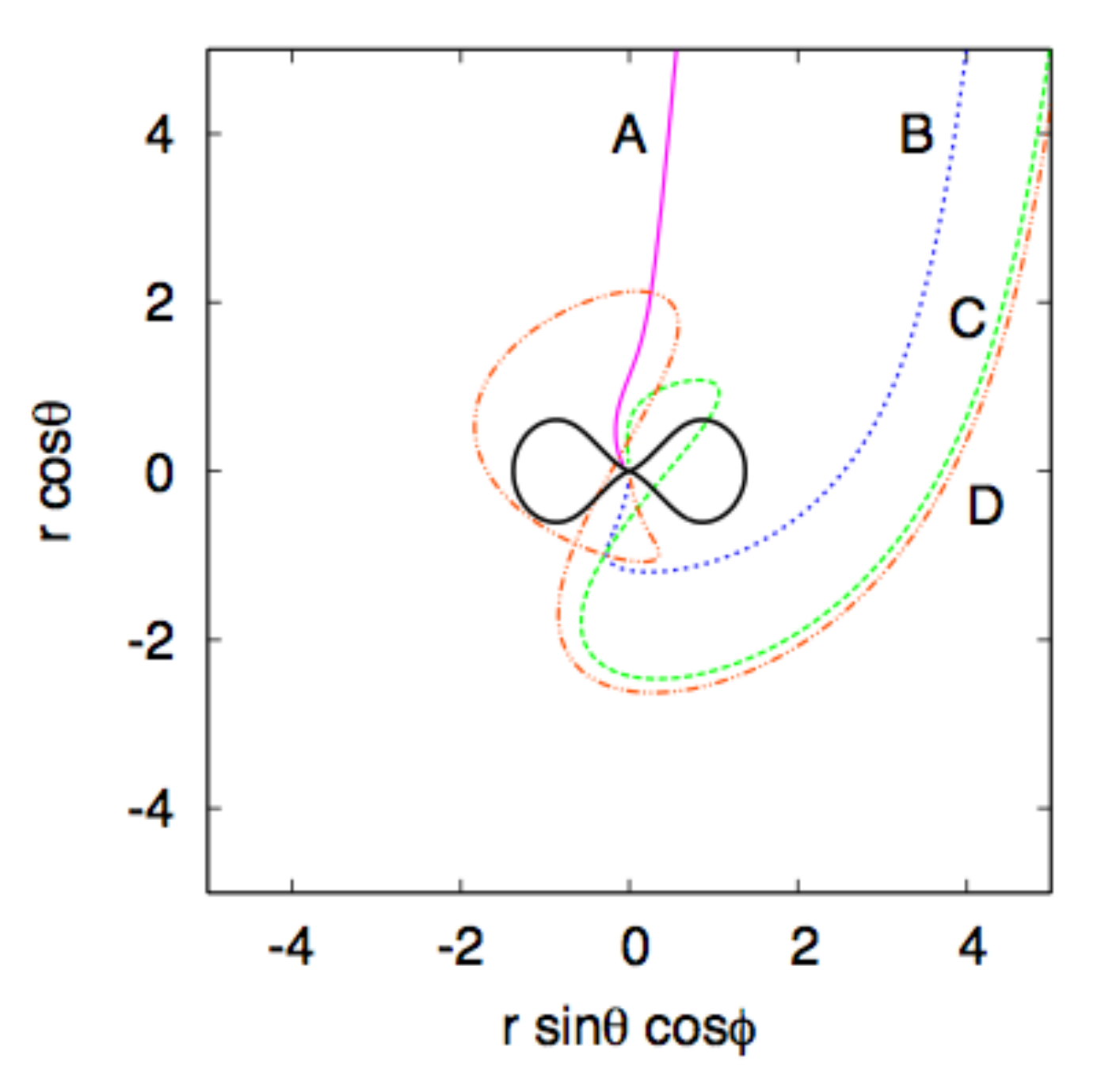}
\includegraphics[type=pdf,ext=.pdf,read=.pdf,width=5.5cm]{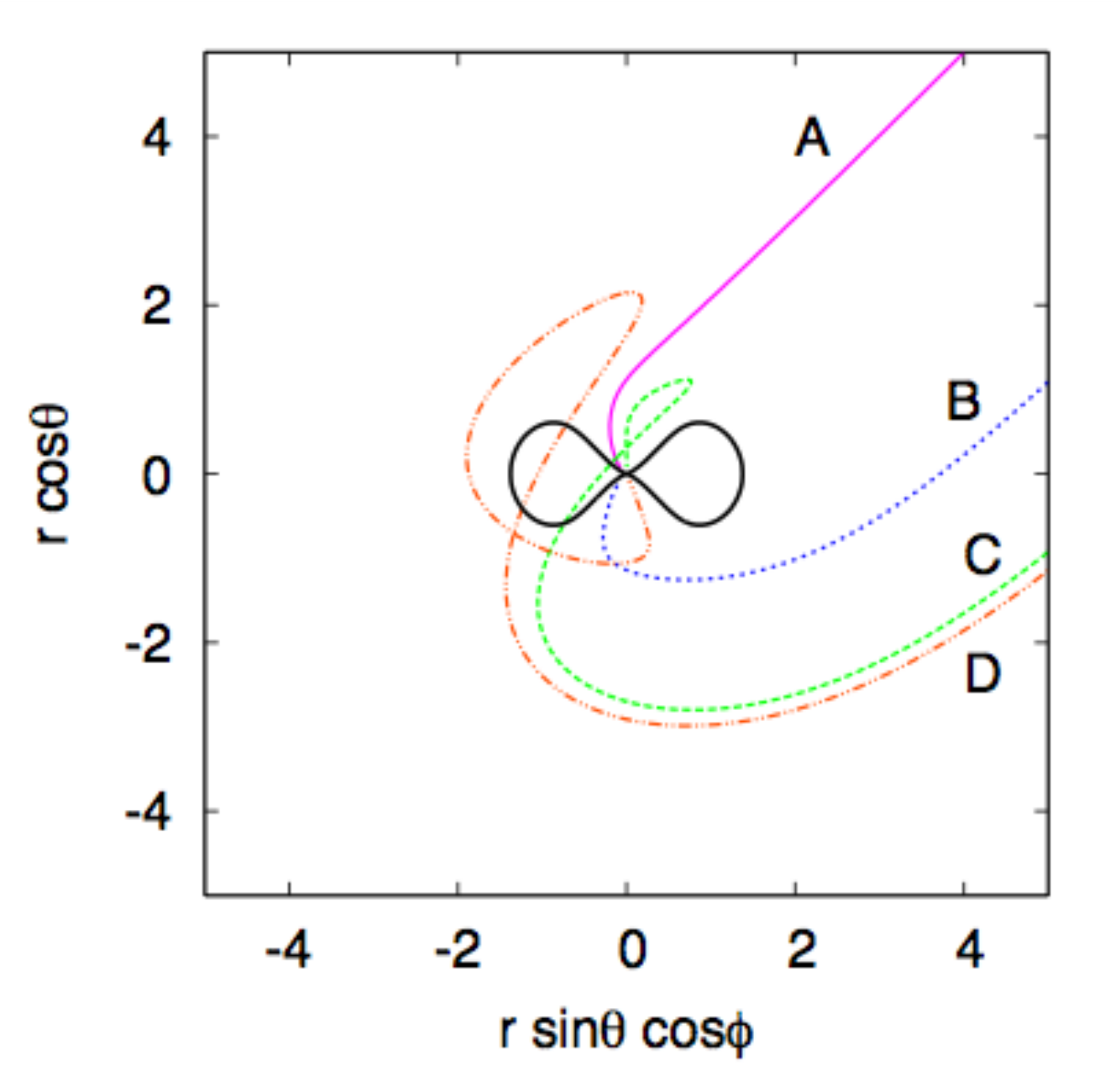}
\includegraphics[type=pdf,ext=.pdf,read=.pdf,width=5.5cm]{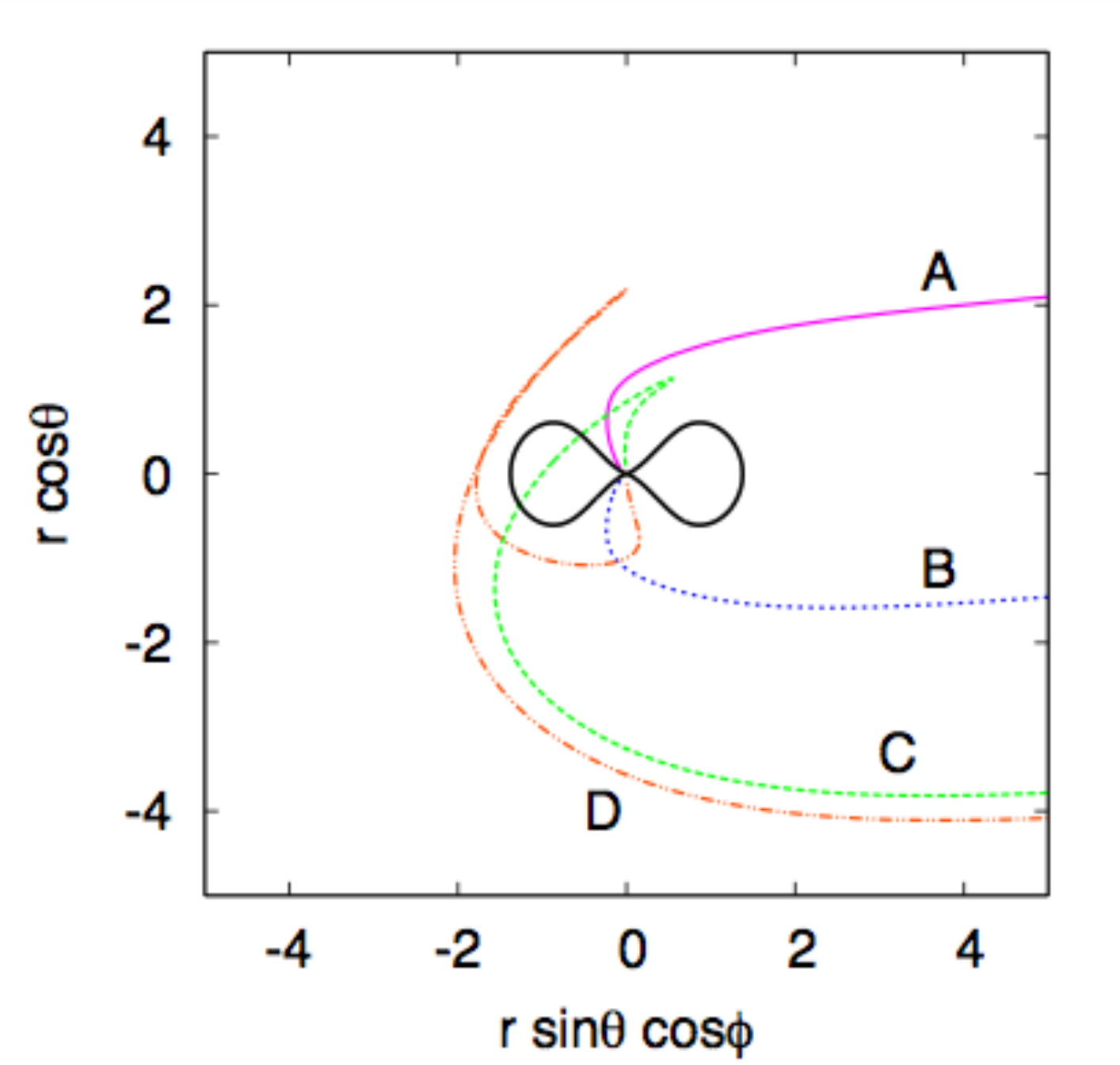}
 \end{center}
\vspace{-0.1cm}
\caption{Trajectories on the $xz$-plane of the light rays A, B, C, 
and D of Fig.~\ref{f1}, for a distant observer with viewing angle 
$i = 5^\circ$ (left panel), $45^\circ$ (central panel), and $85^\circ$ 
(right panel). The primary image of the quantum gravity region is 
formed by light rays like A, while the light rays like B, C, and D 
are responsible for the multiple images. Axes in units $M = 1$.}
\label{f2}
\vspace{0.2cm}
\begin{center}  
\includegraphics[type=pdf,ext=.pdf,read=.pdf,width=5.5cm]{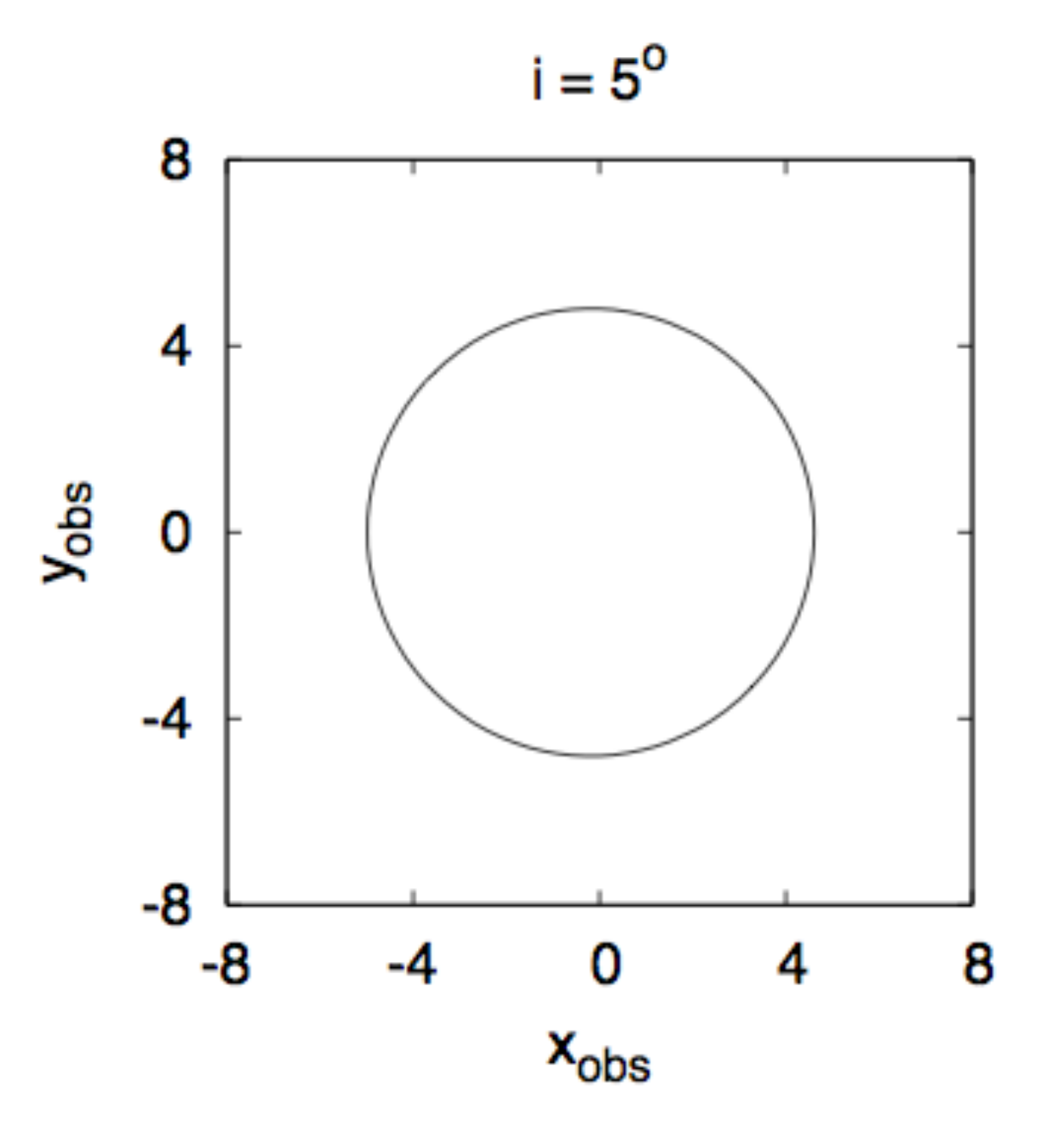}
\includegraphics[type=pdf,ext=.pdf,read=.pdf,width=5.5cm]{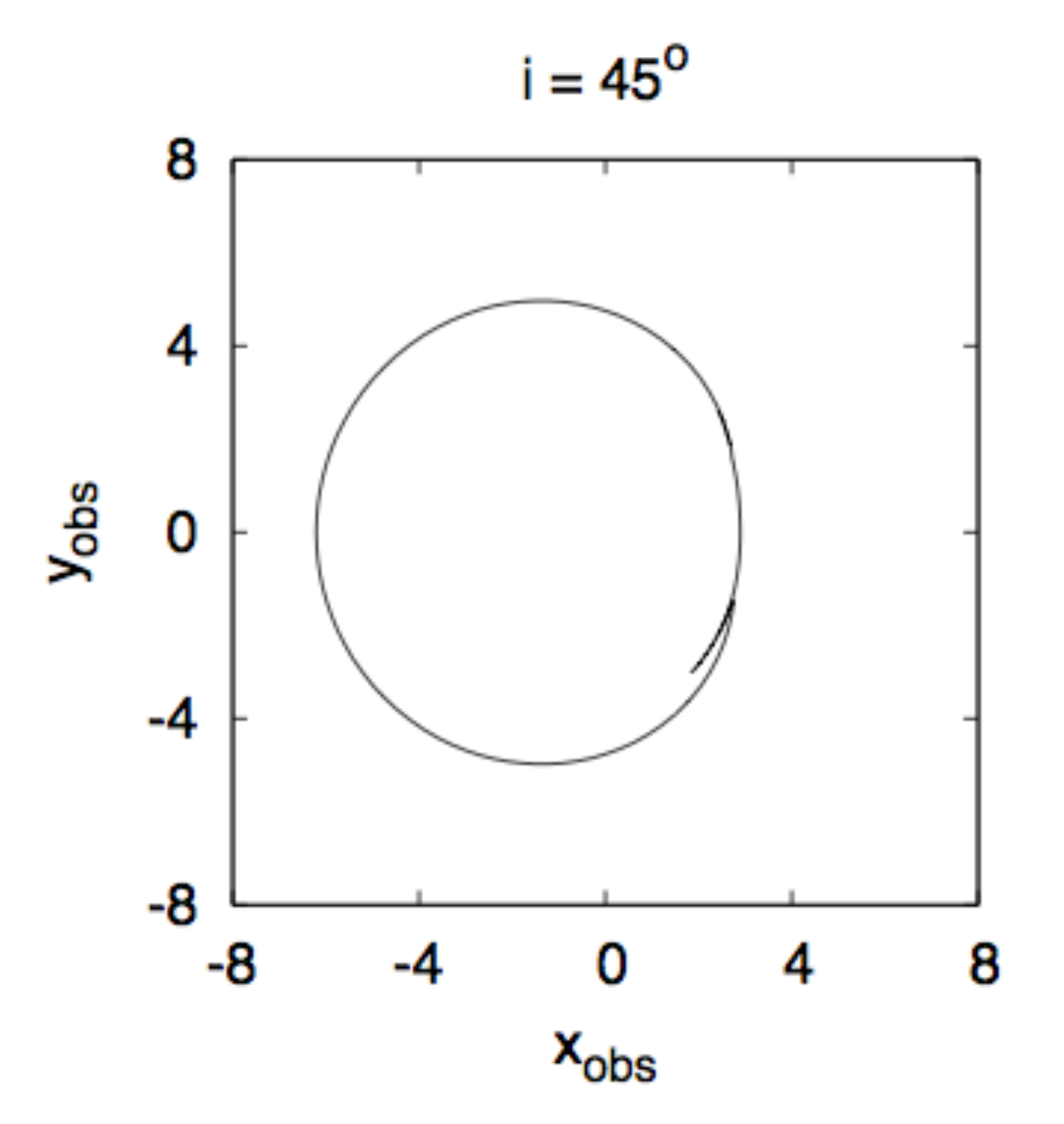}
\includegraphics[type=pdf,ext=.pdf,read=.pdf,width=5.5cm]{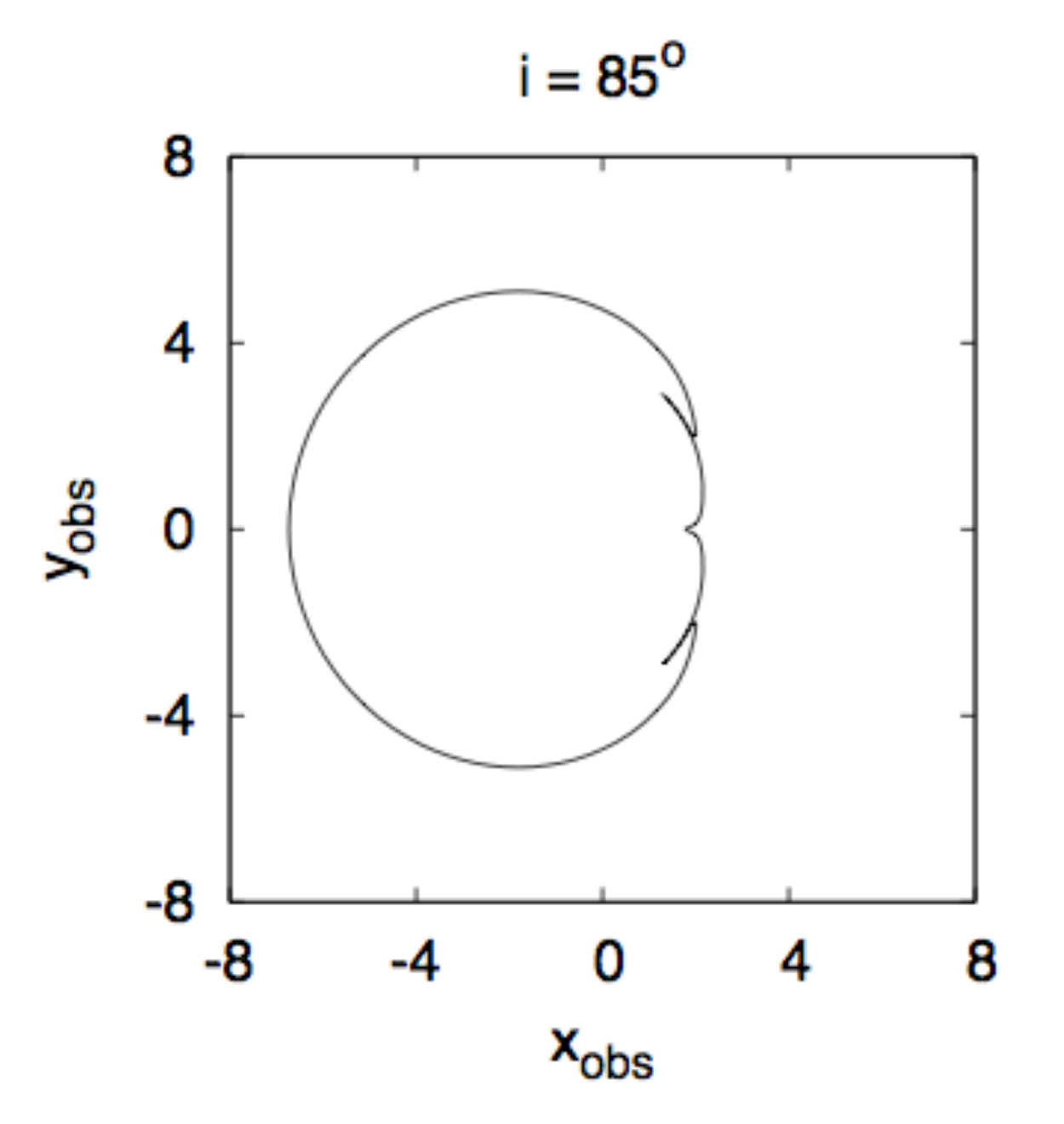}
 \end{center}
 \vspace{-0.3cm}
\caption{As in Fig.~\ref{f1}, for the case of the black hole with 
$a/M = 0.87$ and $\epsilon_3 = 1.0$. Here, the distant 
observer cannot see the quantum gravity region.}
\label{f3}
\end{figure*}

Fig.~\ref{f1} shows the contour of the shadow (black curve) and the 
apparent images of the central singularity (in red) of the first BH
with $a/M = 1.18$ and $\epsilon_3 = -1.0$. The viewing angle of the
distant observer is $i = 5^\circ$ (left panel), $45^\circ$ (central panel), 
and $85^\circ$ (right panel). The BH shadow is computed by 
considering all the photons crossing perpendicularly the screen
of the observer and numerically integrating backward in time the geodesic 
equations for the metric~(\ref{eqm}). The light rays crossing the event 
horizon are inside the shadow, while the ones coming from infinity 
are outside. The resolution of the screen of the observer is $0.001$~$M$. 
More details about the computational procedure can be found in 
previous papers~\cite{shadow}. The quantum gravity region has been 
modeled as a ball of constant radius (in Boyer-Lindquist coordinates) 
$R_{QG} = 0.1$~$M$. The choice of $R_{QG}$ is arbitrary here. 
However, as $R_{QG}$ decreases, the apparent size of the quantum 
gravity region seems to approach a constant value, as shown in 
Tab.~\ref{tab}. {\em Even if the size of the quantum gravity region is 
extremely small, say of order of the Planck scale, the size of its 
apparent image is finite. Such a result is quite interesting and it may
open a new way to probe the Planck scale with astrophysical 
observations}.

Fig.~\ref{f2} shows the trajectories on the $xz$-plane of some light 
rays coming from the quantum gravity region. The primary image is 
formed by light rays traveling along almost-straight paths, while the 
trajectories of the light rays responsible for the other images are 
strongly bent by the gravitational field of the BH.

Fig.~\ref{f3} shows 
the apparent image of the second BH with $a/M = 0.87$ and $\epsilon_3 = 1.0$. 
The shape of this shadow has a very peculiar structure on the side
of the corotating photons ($x_{obs} > 0$ in the central and right panel of 
Fig.~\ref{f3}) due to the existence of two disconnected event horizons; a 
similar feature has never be found for BHs and other compact objects
with trivial topology~\cite{shadow}. It is thus an observational signature
of this BHs with $\epsilon_3 > 0$ and high spin parameter. 
Unlike the images in Fig.~\ref{f1}, here the distant observer cannot 
see the central singularity at $r = 0$, for any value of the viewing
angle. That is due to the presence of the two disconnected horizons, 
above and below the equatorial plane.

\section{Discussion}

As argued in Ref.~\cite{pap1}, rapidly-rotating non-Kerr BHs
may have a topologically non-trivial event horizon. In this work,
we have studied how astrophysical observations can test this
scenario. Interestingly, when these BHs are overspun and the
topology of the horizon changes, the BH central singularity shows
up and, at least in some cases, it can be seen by a distant
observer. In the previous section, we have shown only the
apparent images of two specific cases, but this is enough to 
figure out all the qualitative features. When $\epsilon_3 < 0$
and $a/M > 1$, we can usually see the central region at $r=0$,
but its apparent size decreases/increases for lower/higher
values of $a/M$. For $\epsilon_3 = -1.0$, a BH with
$a/M = 1.18$ is thus the most favorable case in a realistic context, 
because BHs with higher $a/M$ may not be created in Nature.
When $\epsilon_3 > 0$, the central region of the BH 
cannot be seen for spin parameters not exceeding
the equilibrium value of a BH accreting from a thin disk.
For higher values of the spin parameter, the central
region may be seen by a distant observer, but these objects
can unlikely be created, as it seems to occur for a
Kerr naked singularity. On the other hand, the peculiar structure
of the shape of the shadow due to the existence of two
disconnected horizons is common to all these BHs. However,
it is more evident for high values of $a/M$ and $\epsilon_3$.

To test this scenario, it may not be strictly necessary to observe the
exact shape of the BH shadow. When $\epsilon_3 < 0$, the detection 
of the radiation coming from the quantum gravity region may be
enough if it has peculiar properties. Assuming that such
a quantum gravity region can emit some form of radiation, the
latter has likely very high energies, as the central region should
be governed by Planck scale physics. For instance, 
we could try to use these BHs with a visible central singularity
to explain the correlation observed by the Pierre Auger 
experiment between ultra high energy cosmic rays (UHECRs) 
and active galactic nuclei (AGN)~\cite{uhecr}. The observed 
cosmic rays with energies above $6 \cdot 10^{19}$~eV are
definitively difficult to explain with standard acceleration mechanisms,
while here they might be produced by Planck-energy particles 
emitted from the quantum gravity region of AGN, which are indeed 
thought to harbor at their center very rapidly-rotating super-massive 
BHs.

The prediction of the BH apparent image does not require the knowledge
of specific features of the
quantum gravity region. We have just assumed it may emit electromagnetic
radiation. It is however definitively intriguing to think about the
possible properties of this region. For the time being, we do not have
any reliable theory of quantum gravity, and therefore we cannot know
what really happens to the classical singularity. However, 
there are a few scenarios proposed in the literature. For instance,
in the BHs inspired by Loop Quantum Gravity~\cite{loopr}, as well as
in the ones motivated by non-commutative geometries~\cite{kerrr},
the central singularity turns out to be replaced by a Planck length
size region violating the weak energy condition. The size of the
quantum gravity region is therefore independent of the BH mass,
while the energy density changes. At least in the case of the BHs 
in~\cite{kerrr}, the shape of the quantum gravity region is similar
to a ring.

Lastly, let us notice that even in GR, when the space-time has more than 
four dimensions, BHs can have event horizons with topology different 
from the one of the sphere. Thus, non-trivial event horizons might be a 
signature of higher dimensions as well.

\section{Conclusions}

It is thought that the final product of the gravitational collapse
is a Kerr black hole and astronomers have discovered several 
good astrophysical candidates. While there is some indirect 
evidence suggesting that these objects have really an event 
horizon~\cite{horizon}, we do not yet know if the space-time 
around them is described by the Kerr geometry. Recently, 
there has been an increasing interest in the possibility of 
testing the Kerr black hole hypothesis with present and near 
future experiments~\cite{shadow,exp}.

As argued in Ref.~\cite{pap1}, black holes with generic 
deformations from the predictions of General Relativity
may change the topology of 
the horizon above a critical value of the spin parameter. The
accretion process from a thin disk can potentially overspin these
black holes and induce the topology transition, which makes 
the phenomenon astrophysically interesting. In this paper, we
have discussed how such a possibility can be tested. We have 
studied the propagation of light rays in these space-times and 
we have computed the direct image of these objects. Our results
are summarized in Fig.~\ref{f1} (for a black hole more oblate
than a Kerr one) and Fig.~\ref{f3} (for a black hole more prolate 
than a Kerr one). As the contour of the shadow (the black curves 
in Figs.~\ref{f1} and \ref{f3}) is determined by the geometry of the 
space-time, an accurate observation of the direct image of
rapidly-rotating black hole candidates may test this
scenario. Moreover, we have found the more exciting possibility 
that a distant observer may see the central region of these black
holes, where classical physics breaks down and quantum gravity 
effects should appear. The apparent images of this quantum
gravity region, here simply modeled as a luminous ball of constant
radius, are shown in red in Fig.~\ref{f1}.


\begin{acknowledgments}
We would like to thank A. Gnecchi for useful discussions. 
The work of C.B. was supported by Humboldt Foundation.
Research at Perimeter Institute is supported by 
the Government of Canada through Industry Canada and by the Province 
of Ontario through the Ministry of Research \& Innovation. 
\end{acknowledgments}


\end{document}